\documentclass[twocolumn]{openjournal}

\usepackage{orcidlink}
\usepackage{lineno}
\usepackage{comment}
\usepackage{color}
\usepackage{hyperref}
\hypersetup{
	colorlinks=true,
	urlcolor=blue,    
	linkcolor=black,  
	citecolor=blue,   
}
\newcommand{\orcidauthor}[3]{\author{\href{http://orcid.org/#1}{#2$^{#3}$}}}

\usepackage{amsmath}

\newcommand{\Msun}{\ensuremath{\mathrm{M}_\odot}}

\shorttitle{Merger Precursors}
\shortauthors{}

\begin{document}

\title{Merger Precursor: Year-long Transients Preceding Mergers of Low-mass Stripped Stars with Compact Objects\vspace{-1.5cm}}


 \orcidauthor{0000-0002-6347-3089}{Daichi Tsuna}{1,2,*}
 \orcidauthor{0000-0003-2872-5153}{Samantha C. Wu}{1}
 \orcidauthor{0000-0002-4544-0750}{Jim Fuller}{1}
 \orcidauthor{0000-0002-7937-6371}{Yize Dong}{3}
 \orcidauthor{0000-0001-6806-0673}{Anthony L. Piro}{4}
 \affiliation{$^{1}$TAPIR, Mailcode 350-17, California Institute of Technology, Pasadena, CA 91125, USA}
 \affiliation{$^{2}$Research Center for the Early Universe (RESCEU), School of Science, The University of Tokyo, 7-3-1 Hongo, Bunkyo-ku, Tokyo 113-0033, Japan}

\affiliation{$^{3}$Department of Physics and Astronomy, University of California, 1 Shields Avenue, Davis, CA 95616-5270, USA}

\affiliation{$^{4}$Carnegie Observatories, 813 Santa Barbara Street, Pasadena, CA 91101, USA}

\thanks{$^*$E-mail: \href{mailto:tsuna@caltech.edu}{tsuna@caltech.edu}}

\begin{abstract}
Binary mass transfer can occur at high rates due to rapid expansion of the donor's envelope. In the case where mass transfer is unstable, the binary can rapidly shrink its orbit and lead to a merger. In this work we consider the appearance of the system preceding merger, specifically for the case of a low-mass ($\approx 2.5$--$3~M_\odot$) helium star with a neutron star (NS) companion. Modeling the mass transfer history as well as the wind launched by super-Eddington accretion onto the NS, we find that such systems can power slowly rising transients with timescales as long as years, and luminosities of $\sim 10^{40}$--$10^{41}$ erg s$^{-1}$ from optical to UV. The final explosion following the merger (or core-collapse of the helium star in some cases) leads to an interaction-powered transient with properties resembling Type Ibn supernovae (SNe), possibly with a bright early peak powered by shock cooling emission for merger-powered explosions. We apply our model to the Type Ibn SN 2023fyq, that displayed a long-term precursor activity from years before the terminal explosion.
\end{abstract}


\section{Introduction}

\begin{figure*}
    \centering
    \includegraphics[width=0.9\linewidth]{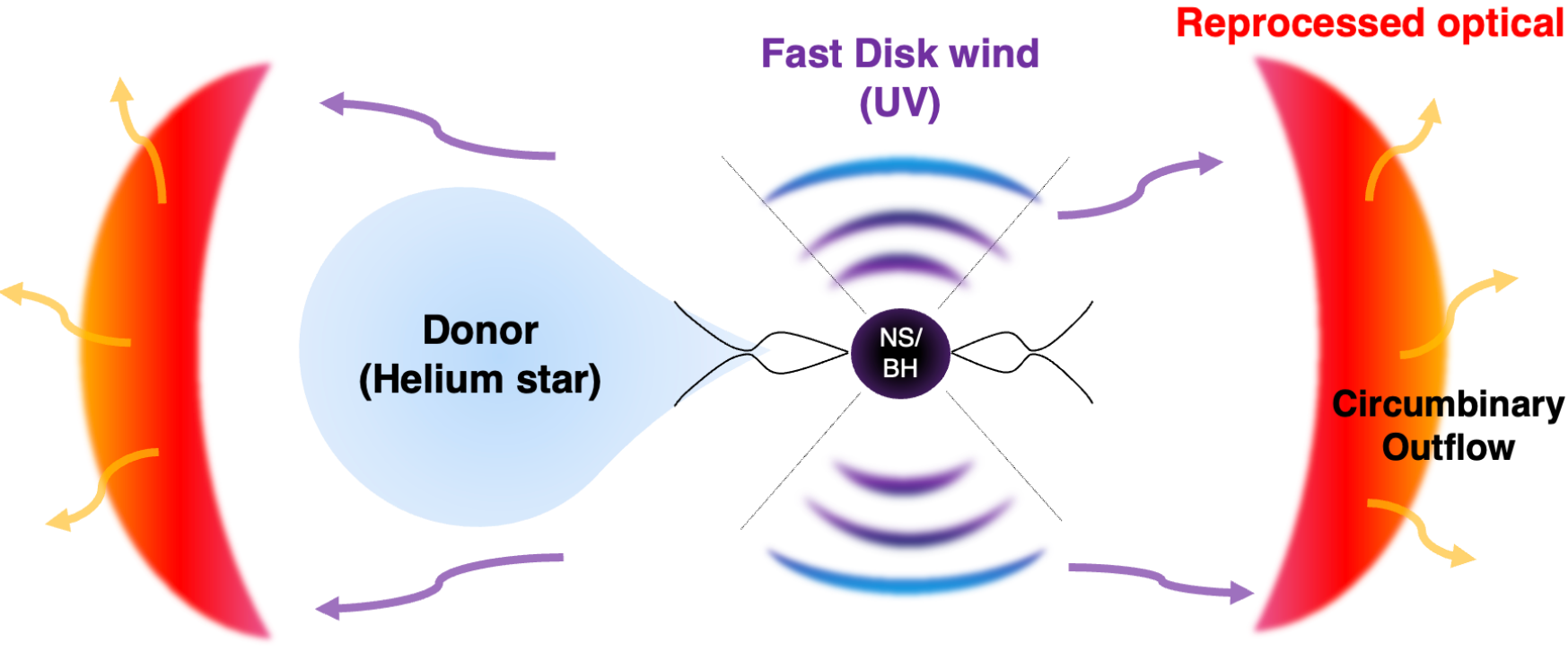}
    \caption{Schematic picture of our merger precursor model (not to scale). Mass transfer at super-Eddington rates from an inflated helium-star donor to a CO companion launches a circumbinary outflow (CBO) through the outer Lagrange point, as well as a fast wind from the CO's accretion disk. These two components power UV emission and reprocessed optical emission with long timescales of up to years.}
    \label{fig:schematic}
\end{figure*}

Precursors of supernovae (SNe) are lumious optical flares observed months to years before the terminal explosion. A growing number of SN precursors have been observed in the previous decade \citep[e.g,][]{Pastorello07,Mauerhan13,Ofek14,Strotjohann21,Jacobson-Galan22,Fransson22,Hiramatsu24,Brennan24,Elias-Rosa24,Dong24}, whose terminal explosions showed signs of dense circumstellar matter (CSM) just outside their progenitors. While the precursor event and the mass loss that creates the dense CSM should be intimately linked \citep[e.g.,][]{Ofek14,Matsumoto22,Tsuna24}, the mechanism for the precursor is not well understood.

A fraction of these are seen in Type Ibn SNe, which are explosions of stripped stars surrounded by dense helium-rich CSM \citep{Pastorello08}. These stars have lost their hydrogen-rich envelope earlier in the evolution, likely due to interaction with a binary companion. It may thus be tempting to consider that binary interactions just prior to the final explosion may be responsible for some of the precursor events and the dense CSM.

Extreme mass loss and luminous emission often co-exist in tight binaries undergoing strong interaction. Powerful, bright outflows due to supercritical accretion onto a compact object (CO) are observed in a Galactic microquasar SS 433 \citep{Margon84,Fabrika04}. More extreme are stellar mergers due to unstable mass transfer, which eject some (or all) of the envelope \citep{Pacynski72,Webbink77,Soker03,Ivanova13} and power transients like luminous red novae \citep[][]{Kulkarni07,Tylenda11,Kochanek14,Blagorodnova17,Pastorello19,Karambelkar23}.

Strong binary interactions can also happen in low-mass helium stars born with masses of $2.5$--$3~M_\odot$, that undergo rapid envelope expansion in the late stages of stellar evolution due to strong helium shell burning \citep{Woosley19,Laplace20,Wu22b}. Recently \cite{Wu22b} showed that binary interactions due to this sudden expansion can lead to unstable mass transfer that triggers mass loss of $0.01$--$1~M_\odot$ in the last years before core-collapse, which reproduce the masses and radii required for the CSM of Type Ibn SNe (Haynie et al. in prep.). 

In this work we consider the appearance of such a helium star and a neutron star (NS) undergoing unstable mass transfer. We construct a physical model of the outflow expected from the binary, and the emission powered by the super-Eddington accretion onto the NS. We find that these systems power luminous emission ($10^{40}$--$10^{41}$ erg s$^{-1}$) from optical to UV, on timescales as long as years until when either the star undergoes core-collapse or the binary merges. The latter case predicts a novel phenomena, ``merger precursors," that can be a new avenue to probe the end stages of stripped stars, as well as the dramatic outcomes of stellar mergers involving COs.

The paper is constructed as follows. In Section \ref{sec:methods} we present our model for the precursor, by formulating the outflows due to the mass transfer process and the emission from these outflows. In Section \ref{sec:results} we show the resulting precursor light curves, as well as the relation of this precursor phenomena to the CSM surrounding the star at the time of its terminal explosion. In Section \ref{sec:2023fyq} we compare our model to observations of SN 2023fyq, a unique Type Ibn SN with a long precursor activity observed for several years. We conclude in Section \ref{sec:conclusion}.

\section{Methods}
\label{sec:methods}

\begin{table*}[]
    \caption{Helium star models at the onset of Roche-Lobe overflow. Model M$x$P$y$ represents a helium star having initial mass (at core hydrogen depletion) $xM_\odot$, put in a binary with a $1.4~M_\odot$ CO companion with initial period of $y$ days. 
    Daggers ($^{\dagger}$) denote the fiducial models we explore the precursor emission in detail, expected to have durations longer than a year.}
    \label{tab:progenitors}
    \centering
    \begin{tabular}{cccccccc}
         \hline
         Model & $R_*$ [$R_\odot$] & $M_*$ [$M_\odot$] & $M_{\rm core}$ [$M_\odot$]\tablenotemark{a}& $q$\tablenotemark{b} & $\gamma_{\rm ad}$ & $\Gamma_s$ & $t_{\rm MT}$ [yr]\tablenotemark{c} \\
         \hline
         M2.467P100$^{\dagger}$ & 62.5 & 1.58 & 1.38 & 0.886 & 1.433 & 1.455 & $\sim 10$ \\
         M2.467P10$^{\dagger}$ & 13.1 & 1.52 & 1.38 & 0.923 & 1.439 & 1.361 & $\sim 40$  \\
         M2.467P1 & 3.6 & 1.44 & 1.37 & 0.969 & 1.428 & 1.288 & $\sim 60$ \\
         M2.754P100$^{\dagger}$ & 61.8 & 2.56 & 1.53& 0.548 & 1.470 & 1.422 & $\sim 0.1$ \\
         M2.754P10 & 11.6 & 1.89 & 1.53 & 0.740 & 1.452 & 1.420 & $\sim 0.3$ \\
         M2.754P1 & 2.5 & 1.72 & 1.52 & 0.812 & 1.446 & 1.319 & $\sim 6$ 
    \end{tabular}
    \tablenotetext{1}{Mass of the carbon-oxygen core.}
    \tablenotetext{2}{Mass ratio $q=M_c/M_*$ for $M_c=1.4~M_\odot$.}
    \tablenotetext{3}{The approximate time until core collapse when mass transfer rates (solved by MESA) exceed $5\times 10^{-4}\, M_{\odot}~\rm{yr}^{-1}$, in years.}
\end{table*}


The schematic picture of our model is depicted in Figure \ref{fig:schematic}. We consider a close ($\lesssim 100~R_\odot$) binary composed of a low-mass helium star and a CO, where the former undergoes rapid expansion in the final years of its life during the oxygen/neon burning phase.

Once the helium star expands, it overfills its Roche lobe and initiates mass transfer onto the CO companion. The mass transfer rate can easily exceed the Eddington rate of the CO, and hence the mass lost from the star almost fully escapes from the binary. A large fraction of the transferred mass is lost through the outer Lagrangian (L2) point \citep{Pejcha16a,Pejcha16b,Macleod18a,Reichardt19,Macleod20,Lu23}, which leaves the system as a circumbinary outflow (CBO). A smaller fraction forms an accretion disk around the CO interior to the CBO, whose accretion rate is still super-Eddington and can launch a radiation-driven disk wind \citep{Lu23,Toyouchi24}.

This mass transfer phase can end in two ways, either by the core-collapse of the helium star or as a merger of the two objects. The latter is triggered by the L2 mass-loss shrinking the orbit, leading to a runaway decrease (increase) in the binary separation (mass-transfer rate) after 10--100 orbital periods \citep[e.g.,][]{Pejcha17,Macleod18a,Macleod20}. After that the star can become tidally disrupted by the CO, and power an explosion driven by extreme accretion onto the CO \citep{Chevalier12,Schroder20,Metzger22}. These two kinds of explosions may be linked to transients with signs of dense CSM, such as Type Ibn or ultra-stripped SNe \citep{Wu22b}.

We propose that during the mass-transfer phase, the disk wind powers luminous emission from optical to UV, lasting as long as years. The luminosity and temperature of the emission depend on the properties of the disk wind and the CBO, and optical emission can be efficiently generated by the denser CBO reprocessing the UV emission from the disk wind. Hereafter we construct a time-dependent emission model starting from realistic helium star models in binaries with a CO companion. While we adopt a NS for the CO, our framework is generally applicable for black hole (BH) companions as well.

\subsection{Initial Binary Models}

As representative cases we adopt two helium stars simulated using MESA \citep{Paxton11,Paxton13,Paxton15,Paxton18,Paxton19,Jermyn23} with the methods in \cite{Wu22b}, which have initial masses of $2.467\ M_\odot$ and $2.754\ M_\odot$ and significantly expand in the final years of their lives. The helium stars are constructed by removing the entire hydrogen envelope after core hydrogen depletion. They are then evolved in a binary with a point-mass companion of $M_c=1.4\ \Msun$ representing a NS, with three initial orbital periods of 100, 10 and 1 day.

Table \ref{tab:progenitors} shows a summary of the properties of the six adopted progenitor models, at the onset of Roche-Lobe overflow during the oxygen/neon burning phase. We also estimate the characteristic adiabatic index $\gamma_{\rm ad}$ and structural index $\Gamma_s$ in the helium layer, which is important for mass transfer and will be used in Section \ref{sec:rlof_model}. We obtain $\gamma_{\rm ad}$ by the mass-weighted mean of the local adiabatic index $\Gamma_1\equiv(d\ln P/d\ln\rho)_{\rm ad}$ over the helium-rich layer of the donor, and $\Gamma_s$ by a power-law fit of the polytropic relation $P\propto \rho^{\Gamma_s}$ in this layer. 

The three initial orbital periods lead to different degrees of envelope stripping throughout its life. The donors with shortest initial periods are subject to more stripping by the companion during helium-shell and core-carbon burning (Case BB mass transfer), which result in smaller masses of the helium layer, smaller radii, and more radiative envelopes with lower $\Gamma_s$.

\subsection{Modeling from Roche-Lobe Overflow to Merger}
\label{sec:rlof_model}

\begin{figure*}
    \centering
    \includegraphics[width=\linewidth]{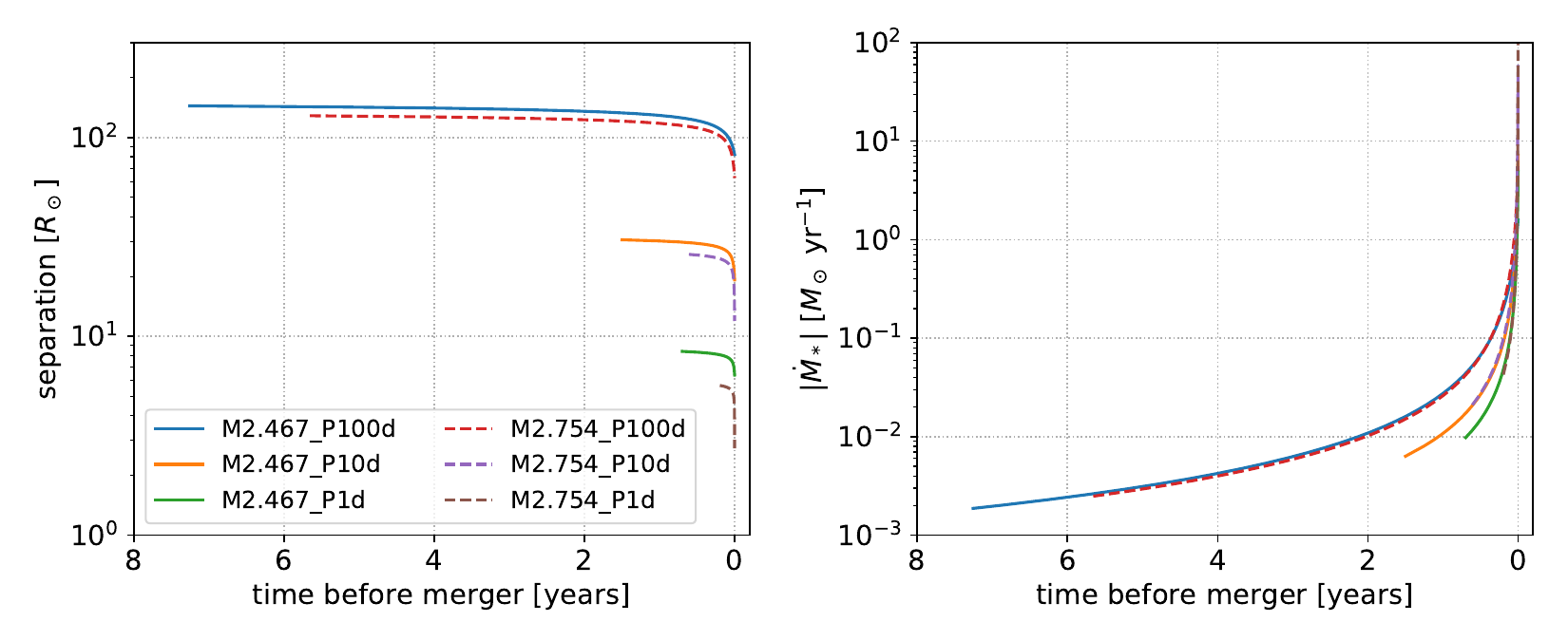}
    \caption{Evolution of the separation and donor mass-loss rate of the six progenitor models. The lines for the mass transfer rates overlap for progenitors with the same initial periods, highlighting the insensitivity of this on the progenitor mass.}
    \label{fig:abin_mdot_vs_t}
\end{figure*}

To follow the mass transfer history from the onset of Roche-Lobe overflow towards merger, we use the python package \texttt{RLOF} \citep{Macleod20,Macleod20_RLOF}. Starting from a binary in a circular orbit with the donor marginally overfilling its Roche Lobe, the code calculates the evolution of separation $a_{\rm bin}$ and donor star mass $M_*$ by solving the following equations (Section 6.2 in \citealt{Macleod20}):
\begin{eqnarray}
\dot{M}_* &=& -\alpha\frac{M_*}{P_{\rm orb}}\left(\frac{R_*-R_L}{R_*}\right)^{(\Gamma_s-1)^{-1}+\frac{3}{2}} \\
\dot{a}_{\rm bin} &=& -2a_{\rm bin}\frac{\dot{M}_*}{M_*}\left[1-\left(\gamma_{\rm loss}+\frac{1}{2}\right)\frac{M_*}{M_* + M_c}\right]
\end{eqnarray}
where $P_{\rm orb}$ is the orbital period, $R_L$ is the (spherically averaged) Roche radius, and $R_*$ is the donor radius. We have simplified the equations from those in \cite{Macleod20} by assuming fully non-conservative mass transfer, i.e. the mass lost from the donor $\dot{M}_*$ fully escapes from the binary without reaching the NS. This is justified by the fact that the typical mass-transfer rate of $\gtrsim 10^{-3}\ \Msun\ {\rm yr}^{-1}$ obtained in our calculations highly exceed the Eddington-limited accretion rate for NSs ($\sim 10^{-8}\ \Msun\ {\rm yr}^{-1}$). 

The dimensionless parameters of order unity $\alpha,\gamma_{\rm loss}$ respectively govern the mass loss and orbital change due to angular momentum loss. To speed up the calculation, fitting formulae for these two parameters were obtained in \cite{Macleod20} with four input parameters, adiabatic index $\gamma_{\rm ad}$, structural index $\Gamma_s$, initial mass ratio $q=M_c/M_*$, and the ratio of stellar spin frequency and orbital frequency $f_{\rm corot}=\Omega_{\rm spin,*}/\Omega_{\rm orb}$. They are
\begin{eqnarray}
    \alpha &\approx& 0.62 [1-0.89(f_{\rm corot}-1)] \nonumber \\
    &&\times \left(\frac{q}{0.1}\right)^{0.68}\left(\frac{\gamma_{\rm ad}}{5/3}\right)^{5.39}\left(\frac{\Gamma_{\rm s}}{5/3}\right)^{-3.25} \\
    \frac{\gamma_{\rm loss}-\gamma_d}{\gamma_{\rm L2}-\gamma_d} &\approx& 0.66 [1-0.30(f_{\rm corot}-1)] \nonumber \\
&&\times \left(\frac{q}{0.1}\right)^{0.08}\left(\frac{\gamma_{\rm ad}}{5/3}\right)^{0.69}\left(\frac{\Gamma_{\rm s}}{5/3}\right)^{-2.17} \label{eq:gamma_frac}
\end{eqnarray}
where $\gamma_d=M_c/M_*$ and $\gamma_{\rm L2}\approx 1.2^2(M_*+M_c)^2/(M_*M_c)$ are the corresponding angular momentum losses if the material escapes from the donor (without influence from the accretor) and from the L2 point respectively. We set the input parameters $(\gamma_{\rm ad}, \Gamma_s, q)$ from our MESA models as in Table \ref{tab:progenitors}, and assume synchronous rotation of the donor with the orbit, i.e. $f_{\rm corot}=1$.

For simplicity the donor radius $R_*$ is assumed to be fixed over the course of mass transfer, which is the default setting of \texttt{RLOF}. This is admittedly a simplifying assumption, and in reality $R_*$ would change in a complex manner as a response to the mass-loss on the donor's dynamical timescale, and by stellar evolution over a comparable timescale as seen in some of the progenitors in \cite{Wu22b} that undergo unstable mass transfer. Adopting an adiabatic response of the donor instead of a fixed $R_*$ can lead towards stabilizing mass transfer for models with large core mass fraction \citep[e.g.,][]{Hjellming87,Soberman97}, although envelope expansion and possibly other non-adiabatic effects like recombination \citep[e.g.,][]{Pavlovskii15} are important for our stars where helium recombines near the surface. In our progenitors, we expect that the rapid envelope expansion by stellar evolution as solved in \cite{Wu22b} initially governs $R_*$, and when the progenitor settles its radius the mass transfer processes solved here would govern the orbital evolution.

With these uncertainties in mind, for all donor models we start the orbital calculations at an initial binary separation of 90\% of the Roche limit \citep{Eggleton83}\footnote{Note that our definition of mass ratio $q$ (same as in \texttt{RLOF}) is the inverse of that in \cite{Eggleton83}.}
\begin{eqnarray}
    a_{\rm bin,0} = 0.9 \left[\frac{0.6+q^{2/3}\ln(1+q^{-1/3})}{0.49}\right]R_*,
\end{eqnarray} 
i.e. the star is overfilling its Roche lobe by $\approx 10\%$. Such overfilling fractions were typically seen in the binary evolution models of \cite{Wu22b}, from the onset of mass transfer until when the progenitor's radius stabilizes. The orbital evolution is terminated either when $a_{\rm bin}=R_*$ or when the entire helium layer is lost, which we define as the onset of the merger.

Figure \ref{fig:abin_mdot_vs_t} shows the evolution of the separation and mass-loss rate $|\dot{M}_*|$ as a function of time. The evolution of $a_{\rm bin}$ ($|\dot{M}_*|$) is qualitatively similar for all models, with a nearly constant (power-law) evolution followed by a runaway decrease (increase). The mass transfer history is largely insensitive to the progenitor, but depends on the initial orbital period that changes the progenitor radius. We find the duration of mass transfer to be from $\sim 0.1$ to $\sim 7$ years.

Whether these stars proceed to merger also depends on the time of the helium star until core-collapse. Table \ref{tab:progenitors} shows the star's remaining lifetime at the onset of mass transfer $t_{\rm MT}$. The lifetime is also sensitive to the helium star's mass, with values ranging from $\sim 0.1$ to $\sim 60$ years. While we calculated until merger for all models assuming the core-collapse has not happened, there can also be cases where mass-transfer (and the precursor) is terminated before merger by core-collapse, especially for heavier progenitors (see also Section \ref{sec:explosion}).

\subsection{Inner Accretion Disk around the Compact Object and the Outer Circumbinary Outflow}
While the above formalism enables time-dependent modeling of the orbit shrinkage and mass transfer, it does not capture the accretion onto the CO and its impact on the fate of the transferred mass. Considering accretion is important for our problem in two aspects. 

First, viscous heating can control the disk accretion rate, by bringing the outer part of the accretion flow to a positive Bernoulli number and driving it away from the system \citep{Narayan95,Lu23}. Second, if the accretion onto the CO is super-Eddington, winds can be launched from the disk as it becomes advection-dominated \citep{Blandford99}. Such disk winds are recently proposed in \cite{Tsuna24} as a powering mechanism for the bright precursors of interacting SNe, and as we show below it is also important for the observational appearance of the systems we model. 

To consider these physics, we use the steady-state one-zone model of \cite{Lu23}, which calculates the fraction $f_{\rm L2}$ of material escaping from L2 by a CBO as a function of $a_{\rm bin}$, $|\dot{M}_*|$ and $q$. The mass outflow rate that leaves the binary as a CBO is
\begin{eqnarray}
    \dot{M}_{\rm CBO} = f_{\rm L2}|\dot{M}_*|,
\end{eqnarray}
and the mass accretion rate of material reaching the inner disk capable of launching a disk wind is then
\begin{eqnarray}
\dot{M}_{\rm acc}=(1-f_{\rm L2})|\dot{M}_*|.
\end{eqnarray}
The value of $f_{\rm L2}$ is determined by the heating-cooling balance in the disk, obeying an alpha-viscosity \citep{Shakura73} with $\alpha=0.1$. We adopt a Rosseland-mean opacity for a helium-dominated composition of solar metallicity (i.e. hydrogen, helium and metal mass fractions of $X=0$, $Y=0.98$, $Z=0.02$ respectively), used in \cite{Lu23} (see their Appendix A).

\begin{figure*}
    \centering
    \includegraphics[width=0.9\linewidth]{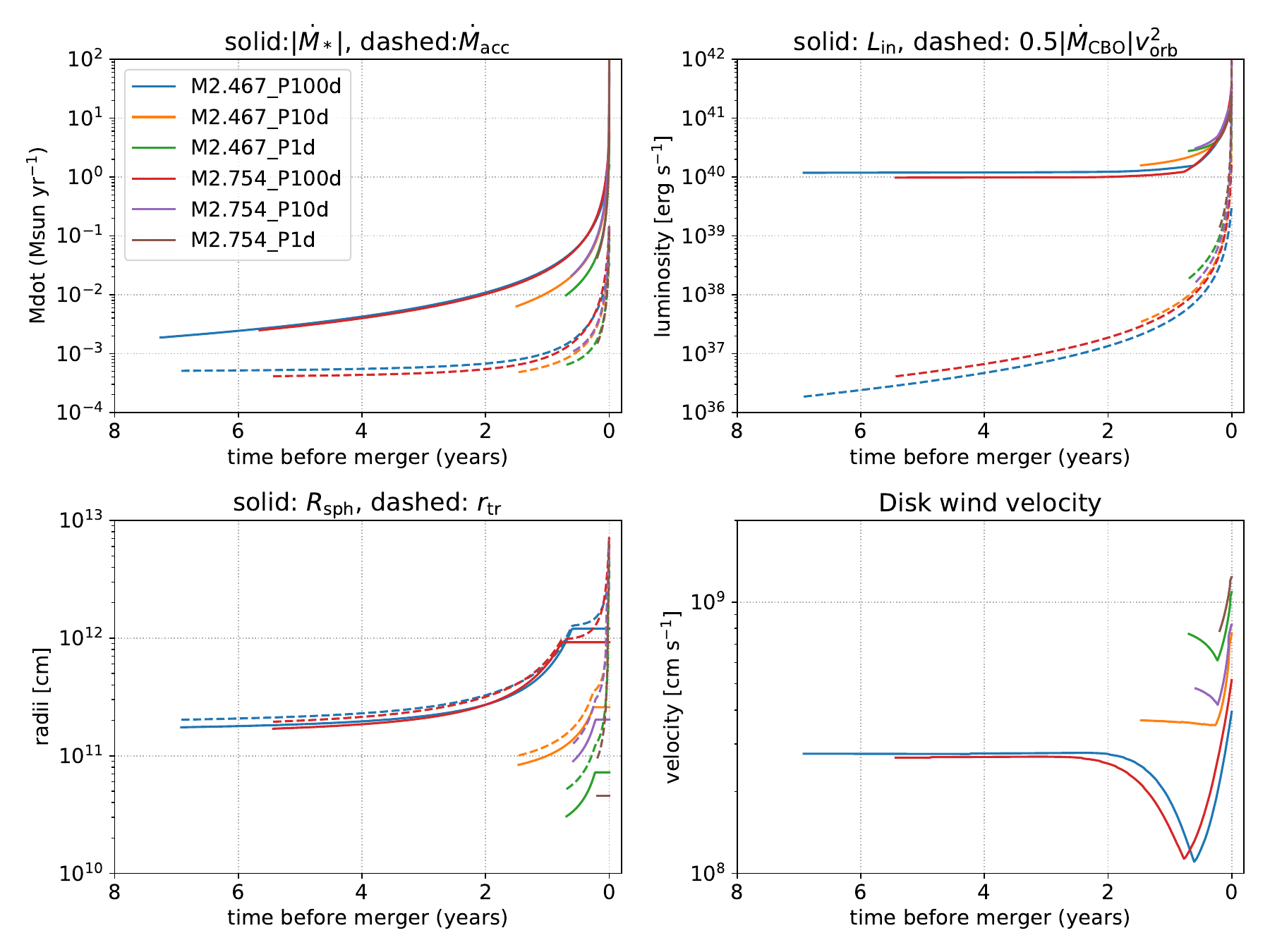}
    \caption{Properties of the disk wind as a function of time. (Top-left): RLOF rate ($\dot{M}_*$) and disk accretion rate ($\dot{M}_{\rm acc}$). (Top-right): Disk input power $L_{\rm in}$ and CBO kinetic luminosity $0.5\dot{M}_{\rm CBO}v_{\rm orb}^2$, where $v_{\rm orb}=\sqrt{G(M_*+M_c)/a_{\rm bin}}$ is the orbital velocity. (Bottom-left): Outermost disk wind radii $R_{\rm sph}$, and trapping radii $r_{\rm tr}$. The kink at $\lesssim 1$ yr before merger corresponds to the transition from $R_{\rm sph}=\kappa_d \dot{M}_{\rm acc}/4\pi c$ to $R_{\rm sph}=R_d$ in equation (\ref{eq:R_out}). (Bottom-right) Asymptotic disk wind velocity $v_{\rm wind}$.}
    \label{fig:disk_wind_properties}
\end{figure*}

In the top-left panel of Figure \ref{fig:disk_wind_properties} we show the time evolution of $|\dot{M}_*|$ and $\dot{M}_{\rm acc}$. The mass-loss from L2 becomes significant ($f_{\rm L2}\approx 1$) at $|\dot{M}_*|\gtrsim 10^{-3}~M_\odot\ {\rm yr}^{-1}$, and at this regime the disk accretion rate is controlled to $\dot{M}_{\rm acc}\lesssim 10^{-3}~M_\odot~{\rm yr}^{-1}$ with the rest escaping as L2 mass loss\footnote{We confirm the internal consistency of this model with the orbital calculations in Section \ref{sec:rlof_model}, as the angular momentum loss prescription used in \texttt{RLOF} is similar to that expected for L2 mass loss (i.e., the fraction in equation \ref{eq:gamma_frac} is close to unity, with a range $0.9$--$1.05$ for our stellar models).}. The model assumes steady state and synchronous rotation of the donor with the orbit. It thus starts to become inaccurate at the ramp-up phase when the donor loses a significant fraction of its envelope within an orbital period and corotation breaks down. This is the case at roughly the final month before merger when $|\dot{M}_*|$ reaches $ 0.1$--$1~M_\odot\ {\rm yr}^{-1}$.

\subsection{Wind Launching from the Inner Accretion Disk}
\label{sec:wind}
The accretion fed to the inner disk is still super-Eddington, and thus we expect fast outflows from disk material advected to inner radii closer to the CO. When the donor is corotating with the orbit, matter flowing from the L1 point contains angular momentum to circularize at a radius \citep{Lu23}
\begin{eqnarray}
    R_{d} &\approx& \frac{(1-x_{\rm L1})^4(1+q)}{q} a_{\rm bin,0} \label{eq:R_d}\\
    x_{\rm L1} &\approx& -0.0355(\log_{10}q)^2 - 0.251\log_{10}q + 0.500
\end{eqnarray}
where $x_{\rm L1}$ is the distance from the donor center to the L1 point, normalized by the binary separation. For simplicity we adopt the initial values for the separation $a_{\rm bin,0}$ and the mass ratio for the entire evolution.

The disk wind outflow is expected be launched within a spherization radius $R_{\rm sph}$ from the CO, where the local accretion luminosity $GM_c\dot{M}_{\rm acc}/R_{\rm sph}$ exceeds the Eddington luminosity $4\pi GM_c c/\kappa_d$ \citep{Begelman79},
\begin{eqnarray}
    R_{\rm sph} = {\min}\left(R_d,\frac{\kappa_d\dot{M}_{\rm acc}}{4\pi c}\right) .
    \label{eq:R_out}
\end{eqnarray}
Here $c$ is the speed of light, and $\kappa_d$ is the opacity in the disk solved in the one-zone model. As typically done when modeling super-Eddington disk wind, we adopt an accretion rate of a power-law function in radius from the CO of $\dot{M}\propto r^{p}$ ($R_{\rm in}<r<R_{\rm sph}$), where a fraction of the accreted mass is blown away as wind at each radii down to the inner edge of the disk $R_{\rm in}$ \citep{Blandford99}. We assume that the velocity of the wind is equivalent to the local escape velocity, carrying a specific energy of $GM_c/2r$. Then the mass outflow rate and kinetic luminosity of the disk wind integrated over the range of $r$ is estimated as
\begin{eqnarray}
    \dot{M}_{\rm wind} &\approx& \left[1-\left(\frac{R_{\rm in}}{R_{\rm sph}}\right)^p\right]\dot{M}_{\rm acc} \\
    L_{\rm wind} &\approx& \frac{p}{2(1-p)}\dot{M}_{\rm acc}\frac{GM_{\rm c}}{R_{\rm in}}\left[\left(\frac{R_{\rm in}}{R_{\rm sph}}\right)^p - \frac{R_{\rm in}}{R_{\rm sph}}\right].
    \label{eq:L_wind}
\end{eqnarray}
For the power-law index $p$ a range of $0.3\leq p \leq 0.8$ is suggested \citep{Yuan&Narayan2014}. In this work we adopt $p=0.5$, and $R_{\rm in}=12$ km appropriate for a NS surface \citep[e.g.,][]{Raaijmakers21}. 

NSs have solid surfaces in contrast to BHs, and we expect the energy advected to the NS to not be swallowed by the NS but be dissipated and re-radiated \citep[e.g.,][]{Narayan95,Ohsuga07}. The energy released by accretion close to the NS is uncertain, and likely depends on the NS's magnetic field and rotation as well as the accretion rate \citep[e.g.,][]{Basko76,Ohsuga07,Piro11,Mushtukov15,King16,Kawashima16,Mushtukov17,Takahashi17}. Here for simplicity we assume that the bound material in the disk which eventually reaches $R_{\rm in}$ carries kinetic and internal energy of total $GM_c/2R_{\rm in}$ (i.e. specific energy of $-GM_c/2R_{\rm in}$). We further assume that this energy is fully dissipated upon reaching the NS surface, and contributes to the energy budget of the disk wind. We thus assume the total input luminosity from the inner disk to be
\begin{eqnarray}
    L_{\rm in} = L_{\rm wind}+\frac{GM_{\rm c}\dot{M}_{\rm acc}}{2R_{\rm in}}\left(\frac{R_{\rm in}}{R_{\rm sph}}\right)^p.
\end{eqnarray}

As shown in the top-right panel of Figure \ref{fig:disk_wind_properties}, the input power from accretion is typically $L_{\rm in}\approx 10^{40}$--$10^{41}\ {\rm erg\ s^{-1}}$ (solid lines), and is initially almost constant due to the time-steady nature of $\dot{M}_{\rm acc}$. They are orders of magnitude larger than the kinetic luminosity of the CBO (dashed lines), whose dissipation by internal shocks have been proposed to power optical/infrared transients preceding stellar mergers \citep[e.g.,][]{Pejcha16a,Pejcha16b}.

\subsection{Emission from the Inner Disk Wind}
We estimate the emission from the disk wind, following the model of wind-reprocessed emission by \cite{Piro20} and \cite{Kremer23} (see also \citealt{Uno20}). Our setup is similar to that depicted in Figure 6 of \cite{Kremer23}, where an optically thick disk wind reprocesses the input power from accretion onto the CO. We slightly update their formalism for our purpose, by adopting a more realistic opacity modeling for accurately obtaining the emission temperature.

For $0<p<1$, the mass budget of the wind mostly comes from the largest radii near $R_{\rm sph}$, while the energy budget is mostly from the smallest radii near $R_{\rm in}$. As the faster wind from inside catches up and collides with the slower wind from outside, we expect the wind power $L_{\rm wind}$ to be dissipated and thermalized near $R_{\rm sph}$, and the merged wind to expand quasi-spherically.

At $r=R_{\rm sph}$ the bulk of the input power $L_{\rm in}$ is carried by radiation. The wind adiabatically expands out to a trapping radius $r_{\rm tr}$, where the diffusion and dynamical times become equal. We solve for $r_{\rm tr}$ by
\begin{eqnarray}
    \frac{\kappa_{\rm tr}\dot{M}_{\rm wind}}{4\pi c} = r_{\rm tr}-R_{\rm sph}
    \label{eq:r_trap_eq}
\end{eqnarray}
where $\kappa_{\rm tr}$ is the opacity at $r_{\rm tr}$ which is determined later, and it is implicitly assumed that $\dot{M}_{\rm wind}$ does not greatly change over the dynamical time. Within $r_{\rm tr}$ the internal energy density scales as $\mathcal{E}(r)\propto r^{-8/3}$ for radiation-dominated gas, and for $r\geq r_{\rm tr}$ the radiative luminosity is approximately constant. The luminosity that diffuses out of the wind is (eq. 10 of \citealt{Piro20})
\begin{eqnarray}
    L_{\rm rad}(r\geq r_{\rm tr}) \approx L_{\rm in}\left(\frac{r_{\rm tr}}{R_{\rm sph}}\right)^{-2/3},
    \label{eq:L_obs}
\end{eqnarray}
where we dropped the time-dependence $dr_{\rm tr}/dt$ from their equation due to the time-steady nature of the system. The asymptotic wind velocity $v_{\rm wind}$ at $r>r_{\rm tr}$ is set by the relation
\begin{eqnarray}
    \frac{\dot{M}_{\rm wind} v_{\rm wind}^2}{2} = L_{\rm in} - L_{\rm rad} = L_{\rm in}\left[1-\left(\frac{r_{\rm tr}}{R_{\rm sph}}\right)^{-2/3}\right].
\end{eqnarray}

At $r\geq r_{\rm tr}$, the density in the wind follows a profile 
\begin{eqnarray}
    \rho(r)=\frac{\dot{M}_{\rm wind}(t_0)}{4\pi r^2 v_{\rm wind}(t_0)}
    \label{eq:rho_CSM}
\end{eqnarray}
where $t_0(r,t)$ denotes the time when the wind shell at $r$ was launched at $r_{\rm tr}$, and is obtained by solving
\begin{eqnarray}
    r = r_{\rm tr}(t_0) +  v_{\rm wind}(t_0)\times (t-t_0).
\end{eqnarray}
We solve the temperature profile in the wind $T(r\geq r_{\rm tr})$ under the flux-limited diffusion approximation,
\begin{eqnarray}
    \frac{\partial T^4}{\partial r} &=& -\frac{1}{\lambda(R)}\frac{\kappa_{\rm R}\rho L_{\rm rad}}{4\pi r^2ac} \label{eq:T_vs_r} \\
    \lambda(R) &=& \frac{2+R}{6+3R+R^2} \label{eq:lmbd_eq}\\
    R &=& \frac{|\partial T^4/\partial r|}{\kappa_{\rm R}\rho T^4} = \frac{L_{\rm rad}}{4\pi r^2acT^4\lambda(R)} \label{eq:R_eq},
\end{eqnarray}
where $\lambda(R)$ is the flux limiter \citep{Levermore81} that takes a value between $1/R$ (optically thin limit; $R\to\infty$) and $1/3$ (optically thick limit; i.e., diffusion approximation), $\kappa_{\rm R}(\rho, T)$ is the Rosseland mean opacity, and $a$ is the radiation constant. The outermost extent of the wind is assumed to obey the  optically thin regime, namely $T^4=L_{\rm rad}/(4\pi r^2 ac)$ and $\partial T^4/\partial r = -2T^4/r$. The equations can then be straightforwardly integrated outside-in, given a model of $\kappa_{\rm R}$. 

We iteratively solve for $\kappa_{\rm tr}$ in equation (\ref{eq:r_trap_eq}) by requiring $\kappa_{\rm tr}=\kappa_{\rm R}(r=r_{\rm tr})$, and obtain $r_{\rm tr}, L_{\rm rad}$ and the density/temperature profiles at $r\geq r_{\rm tr}$. The color temperature of the observed emission is then determined by solving for the color radius $r_{\rm col}$ where the thermalization depth is unity. We solve for $r_{\rm col}$ by
\begin{eqnarray}
    \tau_{\rm eff}(r_{\rm col}) = \int_{r_{\rm col}}^{\infty} \sqrt{3\kappa_{\rm R}\kappa_{\rm abs}}~\rho(r') dr' = 1,
\end{eqnarray}
where $\kappa_{\rm abs}(\rho, T)$ is the absorption opacity which we obtain together with $\kappa_{\rm R}$ as described in Appendix \ref{sec:app_opacity}. Specifically, we adopt opacity tables in order to capture the drop of the opacity at $T\lesssim 10^4$ K due to helium recombination, which would be important for predictions in the optical and near-UV. 

We find that $r_{\rm tr}$ is at most a factor of a few larger than $R_{\rm sph}$, and thus a large fraction of the input power $L_{\rm in}$ can be radiated without adiabatic losses. The wind velocity has a range $v_{\rm wind}\approx 10^8$--$10^9$ cm s$^{-1}$ (bottom-right panel of Figure \ref{fig:disk_wind_properties}), comparable to those of the dense CSM typically inferred in SN Ibn \citep{Pastorello16}.

\subsection{Reprocessing of Disk Wind Emission by the CBO}
When most of the transferred mass is lost by the CBO from the L2 point (i.e. $1-f_{\rm L2}\ll 1$), the CBO outside the binary can dissipate the kinetic energy of the inner disk wind, and also efficiently reprocess the disk wind emission to longer wavelengths. The degree of reprocessing would depend on the density of the CBO, and the fraction of the solid angle subtended by the CBO.

The CBO would have a coverage fraction \citep{Shu79,Pejcha16a,Lu23}
\begin{eqnarray}
    f_\Omega \approx {\rm min}\left(\frac{c_s}{v_{\rm eq}}, 1 \right)
\end{eqnarray}
where $c_s$ is the isothermal sound speed and $v_{\rm eq}$ is the asymptotic speed of the (unbound) CBO \citep{Shu79}. For mergers of low-mass stars, the sound speed in the CBO is much smaller than $v_{\rm eq}$ and this fraction is expected to be typically low ($f_\Omega\lesssim 0.1$; \citealt{Pejcha16a}), except for very close to merger \citep{Pejcha16b,Macleod18b}. However, in our case with a CO we expect that the CBO can be ionized and heated by photons from the inner disk wind \citep{Lu23}.

In the test-particle limit, the CBO is unbound from the binary for mass ratios of $0.064<q<0.78$ \citep{Shu79}, with asymptotic speeds dependent on $q$ (their Table 2; see also Figure 3 of \citealt{Pejcha16a}) of
\begin{eqnarray}
    v_{\rm eq}&\lesssim& 0.3\sqrt{G(M_*+M_c)/a_{\rm bin}} \nonumber\\
    &\sim& 20\ {\rm km\ s^{-1}}\left(\frac{M_*+M_c}{3.5~M_\odot}\right)^{1/2}\left(\frac{a_{\rm bin}}{100~R_\odot}\right)^{-1/2}.
\end{eqnarray}

The density in the CBO is
\begin{eqnarray}
    \rho(r) = \frac{\dot{M}_{\rm CBO}}{4\pi r^2 v_{\rm eq}f_\Omega},
\end{eqnarray}
and the scattering optical depth of the CBO in the vertical direction is independent of $f_\Omega$ and follows
\begin{eqnarray}
    \tau_{\rm scat}(r)&\approx& \frac{\kappa_{\rm scat}\dot{M}_{\rm CBO}}{4\pi r v_{\rm eq}} \nonumber \\
    &\sim& 70 \left(\frac{\kappa_{\rm scat}}{0.2\ {\rm cm^2\ g^{-1}}}\right)\left(\frac{\dot{M}_{\rm CBO}}{10^{-3}M_\odot\ {\rm yr}^{-1}}\right) \nonumber \\
    &&\times \left(\frac{r}{100R_\odot}\right)^{-1}\left(\frac{v_{\rm eq}}{20~{\rm km\ s^{-1}}}\right)^{-1}.
\end{eqnarray}
While the CBO is optically thick ($\tau_{\rm scat}\gg 1$) close to the binary, $c/\tau_{\rm scat}\gg v_{\rm eq}$ and radiative diffusion is much faster than the adiabatic expansion. Hence we can crudely estimate the sound speed by assuming that the temperature of the CBO is set by balance between radiative cooling and heating by wind emission. Solving for the temperature by equating the radiative cooling rate $4\pi r^2 acT^4/\tau_{\rm scat}$ \citep{Lu23} and the heating rate $f_\Omega L_{\rm rad}$, we obtain the isothermal sound speed
\begin{eqnarray}
    c_s &\approx& \sqrt{\frac{k_BT}{\mu m_p}} \nonumber \\
    &\sim& 20\ {\rm km\ s^{-1}} \nonumber \\
    &&\times \mu^{-1/2} f_\Omega^{1/8}  \left(\frac{L_{\rm rad}}{10^{40}~{\rm erg\ s^{-1}}}\right)^{1/8} \left(\frac{\dot{M}_{\rm CBO}}{10^{-3}M_\odot\ {\rm yr}^{-1}}\right)^{1/8} \nonumber\\
    &&\left(\frac{\kappa_{\rm scat}}{0.2\ {\rm cm^2\ g^{-1}}}\right)^{1/8}  \left(\frac{r}{100R_\odot}\right)^{-3/8}\left(\frac{v_{\rm eq}}{20~{\rm km\ s^{-1}}}\right)^{-1/8}
\end{eqnarray}
where the mean molecular weight $\mu\approx 2\ (1.3)$ for singly (doubly) ionized helium.
These estimates imply that depending on the mass ratio and separation, the resulting $f_{\Omega}$ can be close to unity in the vicinity of the binary. However our estimates do not take into account the complicated thermal processes near the L2 point and in the CBO, as well as the possible effects of finite gas pressure \citep{Pejcha16b} or ram pressure/shock heating by our disk wind neglected in the test-particle limit.

Given the uncertainties, in this work we adopt $f_\Omega(\leq 1)$ as a free parameter of the model, and assume that it is independent of radius and time for simplicity. We assume that the CBO is emitted from $a_{\rm bin}$ with a latitude-dependent density profile
\begin{eqnarray}
    \rho_{\rm CBO}(r,\theta) = 
    \begin{cases}
       f_{\rm L2}|\dot{M}_*|/(4\pi f_\Omega r^2 v_{\rm eq}) & (|\sin\theta|\leq f_\Omega) \\
       0 & (|\sin\theta|>f_\Omega)
    \end{cases}
    \label{eq:MdotCBO}
\end{eqnarray}
where $\theta$ ($-\pi/2\leq \theta \leq \pi/2$) is the latitude with $\theta=0$ set to the orbital plane. As in the previous section the disk wind is launched isotropically with mass-loss rate $\dot{M}_{\rm wind}$.
Thus a fraction $1-f_\Omega$ is subtended by the unimpeded disk wind, whereas in the other $f_\Omega$ the wind and its emission are affected by the CBO.

We also consider the collision between the disk wind and the CBO, which is important for the emission when $f_\Omega$ is close to unity. For latitudes of $|\sin\theta|\leq f_\Omega$, the disk wind collides with the CBO, and the kinetic energy of the wind is converted to radiation via shocks. We assume that the collision merges the disk wind and the CBO, creating a momentum conserving outflow with mass-loss rate $\dot{M}_{\rm out}=\dot{M}_{\rm CBO}+f_\Omega\dot{M}_{\rm wind}$ and velocity
\begin{eqnarray}
    v_{\rm out} &\approx& \frac{\dot{M}_{\rm CBO}v_{\rm eq}+f_\Omega\dot{M}_{\rm wind} v_{\rm wind}}{\dot{M}_{\rm CBO}+f_\Omega\dot{M}_{\rm wind}},
\end{eqnarray}
where we adopt $v_{\rm eq}=0.3v_{\rm orb}$. From energy conservation, the internal energy generated by the dissipation of the wind energy is
\begin{eqnarray}
    L_{\rm sh} &=& \frac{1}{2}f_\Omega\dot{M}_{\rm wind} v_{\rm wind}^2 + \frac{1}{2}\dot{M}_{\rm CBO}v_{\rm eq}^2 \nonumber \\
    &&-\frac{1}{2}(\dot{M}_{\rm CBO}+f_\Omega\dot{M}_{\rm wind})v_{\rm out}^2 \nonumber \\
    &=& \frac{1}{2} \frac{f_\Omega\dot{M}_{\rm wind}\dot{M}_{\rm CBO}}{f_\Omega\dot{M}_{\rm wind}+\dot{M}_{\rm CBO}}(v_{\rm wind}-v_{\rm eq})^2.
\end{eqnarray}
Under this assumption, the situation converges into a wind-reprocessing problem similar as in the previous section. The input luminosity is modified to $f_{\Omega}L_{\rm rad} + L_{\rm sh}$ at $r\approx a_{\rm bin}$, and the density at $r\gtrsim a_{\rm bin}$ becomes larger than that expected from the disk wind by a factor
\begin{eqnarray}
 &&\frac{(\dot{M}_{\rm CBO}+f_\Omega\dot{M}_{\rm wind})/v_{\rm out}}{\dot{M}_{\rm wind}/v_{\rm wind}} \nonumber \\
 &=& \left(1+\frac{\dot{M}_{\rm CBO}}{f_\Omega\dot{M}_{\rm wind}}\right)^2 \left(1+\frac{\dot{M}_{\rm CBO}v_{\rm eq}}{f_{\Omega}\dot{M}_{\rm wind}v_{\rm wind}}\right)^{-1}.
\end{eqnarray}
In the typical case of $f_{\rm L2}\approx 1$ and wind dominating the momentum over the CBO, the density is enhanced from the disk wind by a factor $\approx f_{\Omega}^{-2}[f_{\rm L2}/(1-f_{\rm L2})]^2\gg 1$. For example, for the M2.467\_P100d model with the longest evolution, the value of $f_{\rm L2}$ evolves initially from $f_{\rm L2}\approx 73\%$ to $f_{\rm L2}\approx 96\%$ at one year before merger. This enhancement factor thus evolves by two orders of magnitude from $\approx 7f_\Omega^{-2}$ to $\approx 700f_\Omega^{-2}$, making reprocessing substantially more efficient with time.

We obtain the luminosity and color temperature of the emission reprocessed by this merged outflow, through repeating the same wind-reprocessing calculations by updating the input parameters
\begin{eqnarray}
     && (L_{\rm in},~ \dot{M}_{\rm wind},~ R_{\rm sph}) \nonumber \\
    &\to& (f_{\Omega}L_{\rm rad} + L_{\rm sh},~ \dot{M}_{\rm out},~ a_{\rm bin}).
\end{eqnarray}

For the present case, we find that as the binary approaches merger the mass outflow rate $\dot{M}_{\rm out}$ rapidly increases over the dynamical time $\approx (r_{\rm tr}-a_{\rm bin})/v_{\rm out}$. This invalidates the steady-state approach to estimate $r_{\rm tr}$ and the observed luminosity, done in equations (\ref{eq:r_trap_eq}) and (\ref{eq:L_obs}). We thus do not calculate the emission when $\dot{M}_{\rm out}$ increases at a timescale much faster than the dynamical time at $r_{\rm tr}$, which is set by a simple criterion
\begin{eqnarray}
    \frac{\dot{M}_{\rm out}}{d\dot{M}_{\rm out}/dt} < \frac{r_{\rm tr}-a_{\rm bin}}{v_{\rm out}}.
\end{eqnarray}
In our binary models, this condition is typically met in the last month before merger for the 100 day models, and in the last few weeks for the 10 day models. We expect the light curve afterwards to continue rising, and be eventually overtaken by the terminal explosion (if any) due to merger (see Section \ref{sec:explosion}). Radiation hydrodynamical calculations would be necessary to accurately predict both luminosity and temperature at this phase\footnote{The outflow from the binary can significantly change its direction and opening angle as mass-loss undergoes a runaway \citep[e.g.,][]{Morris06,Pejcha16b,Macleod18b,Reichardt19}, which may also invalidate the assumption of constant $f_\Omega$ around this regime.}.

\section{Results}
\label{sec:results}
\subsection{Precursor Optical/UV Light Curves}
\begin{figure}
    \centering
    \includegraphics[width=\linewidth]{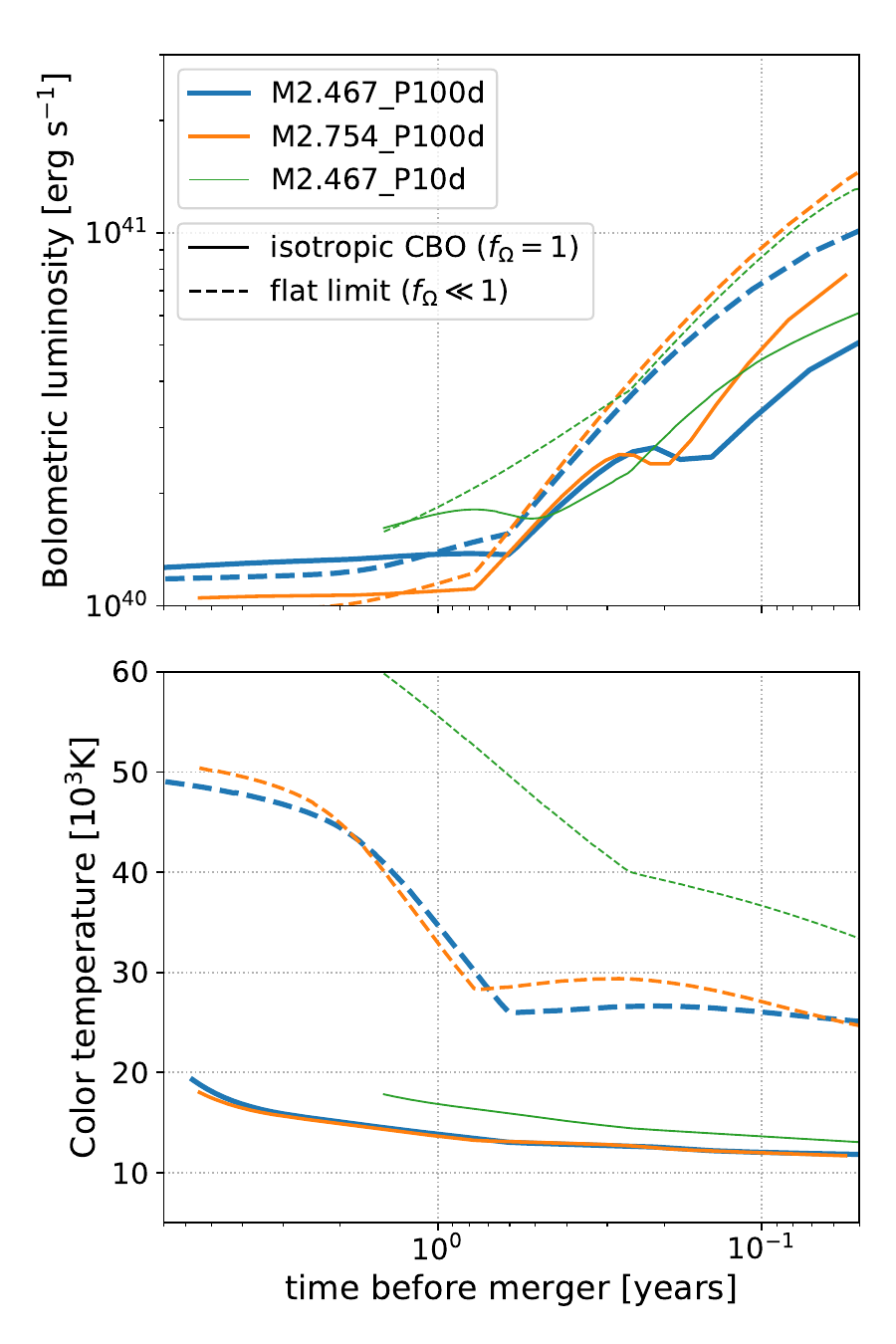}
    \caption{Expected luminosity and temperature of the precursor, for three models and two limiting cases of the CBO's structure, with covering fraction $f_\Omega=1$ (spherical CBO; solid lines) and $f_\Omega\ll 1$ (flat CBO, with no reprocessing of inner disk wind emission; dashed lines). Note that some of our $f_\Omega=1$ models are truncated several weeks before merger when our assumption of steady state breaks down.}
    \label{fig:lum_temp_withcbo}
\end{figure}

Figure \ref{fig:lum_temp_withcbo} shows the luminosity and color temperature of the precursor emission, focusing on three binary models in Table \ref{tab:progenitors} that show long-term ($\gtrsim 1$ yr) precursors of our interest. We consider two limiting cases of $f_\Omega$, that sets the structure of the CBO. The first is $f_\Omega=1$, which corresponds to an isotropic, spherical CBO with largest reprocessing of the emission from the inner disk wind. The second is $f_\Omega\ll 1$ which corresponds to a flat CBO with negligible reprocessing, i.e. we calculate emission from the disk wind only. 

In the absence of reprocessing by the CBO ($f_\Omega\ll 1$, dashed lines), we find luminosities of $10^{40}$--$10^{41}$ erg s$^{-1}$ and color temperatures of $T_{\rm col}\gtrsim$ 2$\times 10^4$ K, i.e. mainly emitting in far-UV. The kinks at $\lesssim 1$ yr before merger reflect the transition of the spherization radius with $\dot{M}_{\rm acc}$ from $R_{\rm sph}=\kappa \dot{M}_{\rm acc}/4\pi c$ to $R_{\rm sph}=R_d$ in equation (\ref{eq:R_out}), which alters the time dependence of $L_{\rm in}$.

The $f_\Omega=1$ model with maximum CBO reprocessing (solid lines) has a temperature reduced to as low as $\sim 10^4$ K, and is brighter in optical than the incident disk wind emission. In the final several months the luminosity for $f_\Omega=1$ becomes lower than that of $f_\Omega\ll 1$ by a factor $\lesssim 2$, as there is additional photon trapping by the CBO that reduces the luminosity due to adiabatic expansion. This photon trapping effect also creates a dip in the light curve around a few months before merger. We find this to be due to the temporary enhancement of $\kappa_{\rm tr}$ as the temperature at $r_{\rm tr}$ crosses (a few) $\times 10^5$ K, where the large number of bound-bound transitions by iron increases the opacity (see also Appendix \ref{sec:app_opacity}).

\begin{figure}
    \centering
    \includegraphics[width=\linewidth]{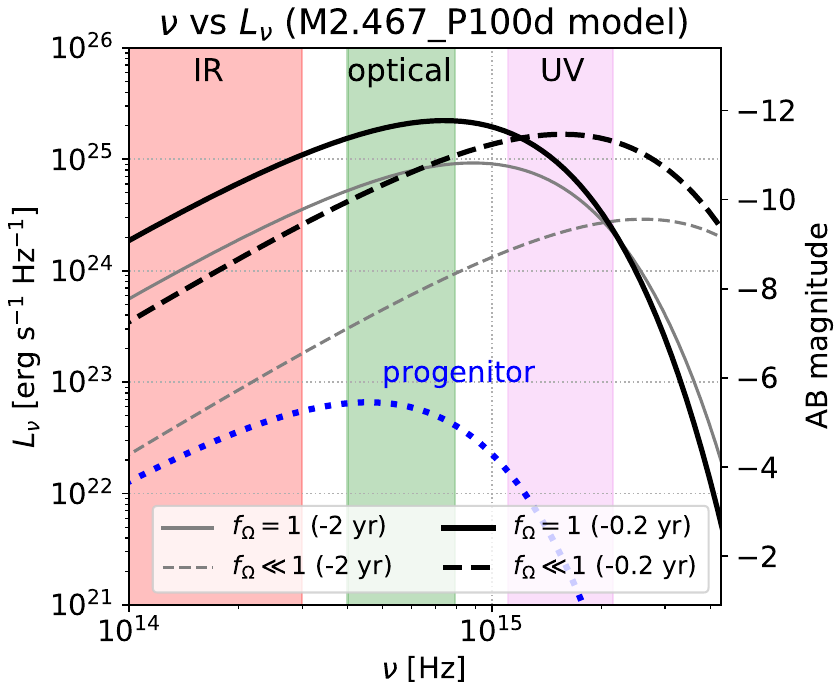}
    \caption{Expected SED of the precursor for the M2.467P100d model at 2 and 0.2 years  before merger, shown in gray and black respectively. The solid and dashed lines are for the two cases for the CBO's structure, as in Figure \ref{fig:lum_temp_withcbo}. The dotted line shows the SED of the progenitor He star.}
    \label{fig:precursor_spec}
\end{figure}

This dichotomy in the temperatures of the two limits indicate that depending on $f_\Omega$, the source can be bright not only in UV but also in optical. Thus optical and UV surveys are complementary for finding these precursors. To see this more quantitatively, we estimate the spectral energy distribution (SED) assuming it follows a greybody of temperature $T_{\rm col}$
\begin{eqnarray}
    L_\nu &=& 4\pi R_{\rm gr}^2\frac{2\pi h\nu^3}{c^2}\frac{1}{\exp[h\nu /k_BT_{\rm col}]-1}
\end{eqnarray}
where $R_{\rm gr}=(L_{\rm rad}/4\pi \sigma_{\rm SB}T_{\rm col}^4)^{1/2}$ and $\sigma_{\rm SB}$ is the Stefan-Boltzmann constant. We also calculate from $L_\nu$ the AB magnitude
\begin{eqnarray}
    M_{\rm AB}= -2.5\log_{10}\left[\frac{L_\nu/4\pi (10~{\rm pc})^2}{3631\ {\rm Jy}}\right].
\end{eqnarray}
As an example, Figure \ref{fig:precursor_spec} shows the optical-UV SED of the precursor, for the two limiting cases of $f_\Omega$. Differences in $L_\nu$ in the UV and optical for the two cases can be as large as an order of magnitude. 

In the last months before merger, for $f_\Omega=1$ ($f_\Omega\ll 1$) we expect the source to have AB magnitudes of -11 to -12 mag in the optical (UV), and about 1--2 mag dimmer in the UV (optical). This is similar in luminosity to the dimmest precursors previously reported, in e.g. SN 2019uo \citep{Strotjohann21} and SN 2020tlf \citep{Jacobson-Galan22}. They are promising targets for future optical and UV surveys such as Rubin \citep{Ivezic19}, ULTRASAT \citep{Ben-Ami22}, and UVEX \citep{Kulkarni21}, in addition to pre-SN outbursts that are predicted to be detectable by Rubin \citep{Tsuna23,Strotjohann24}. For single-visit depths of 24--25 mag for the more sensitive Rubin/UVEX, we expect the precursors in our model to be detectable from months before merger for sources out to $\sim 100$ Mpc.

We finally note that for $f_\Omega<1$, a fraction of the emission from the wind termination shock may escape without being reprocessed by the CBO. Given the wind velocities of $v_{\rm wind}=10^3$--$10^4$ km s$^{-1}$, heating by such shocks may furthermore produce an X-ray counterpart. The large column density of the CBO likely requires a line of sight away from the CBO to observe this.

\subsection{Density Profile of the Circumstellar Matter}
Our model also gives a time-dependent density profile of the CSM, similar to equation (\ref{eq:rho_CSM}) but for the merged outflow (CBO + disk wind), as 
\begin{eqnarray}
    \rho(r,t)=\frac{\dot{M}_{\rm out}(t_0)}{4\pi r^2 v_{\rm out,\infty}(t_0)},
    \label{eq:rho_CSM_CBO}
\end{eqnarray}
where $v_{\rm out,\infty}$ is the asymptotic velocity of the outflow at the trapping radius, and $t_0$ solved from
\begin{eqnarray}
    r = a_{\rm bin}(t_0) +  v_{\rm out,\infty}(t_0)\times (t-t_0).
\end{eqnarray}

\begin{figure}
    \centering
    \includegraphics[width=\linewidth]{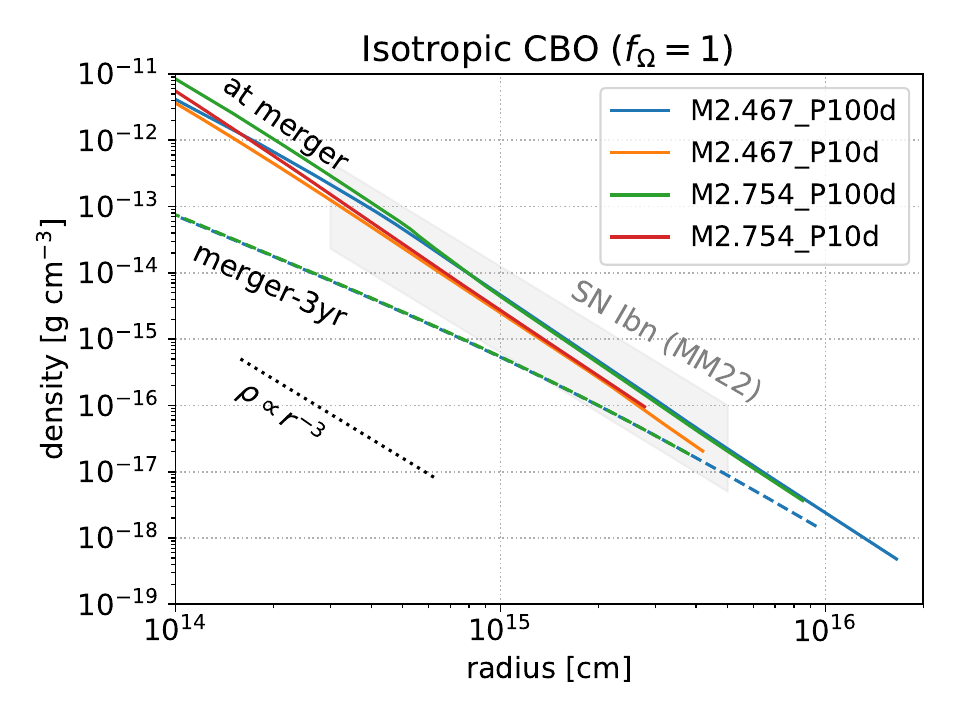}
    \caption{Expected density profile of the extended CSM for an isotropic outflow of $f_\Omega=1$. Solid lines show the profiles at the time of merger, while dashed lines show those at 3 years before merger for the two P100d models. The gray shaded region shows the density profile inferred from light curve modeling of Type Ibn SNe \citep{Maeda22}.}
    \label{fig:CSM_profile}
\end{figure}

We show these density profiles at the onset of merger in Figure \ref{fig:CSM_profile}, for the case of spherical symmetry with $f_\Omega=1$. The outer profile at $r\gtrsim 10^{14}$ cm deviates from a canonical wind profile of $\rho\propto r^{-2}$, expected for a constant mass loss rate and wind velocity. This is due to the significant change in the mass-loss rate over the time the CSM expands at this radii. As material ejected earlier has lower mass-loss rates, the profile becomes steeper than the wind profile. 

This time-dependence of the mass-loss rate also creates a time-dependence on the CSM profile, even for the same progenitor model. We overplot in Figure \ref{fig:CSM_profile} as dashed lines the profiles at 3 years before merger, for the two stellar models with initial periods of 100 days. Closer to merger the density at inner radii is highly enhanced, due to the much higher rate of the previous mass loss episode responsible for the CSM at those radii. Thus, if the progenitor exploded before the merger time defined here (e.g. as a core-collapse SN), it would encounter less dense CSM at $r\lesssim 10^{15}$ cm than if it exploded close to or via merger.

We overall find a profile close to $\rho\propto r^{-3}$ in the outer part of the CSM around $10^{15}$ cm, which is responsible for powering the emission around the optical peak for Type Ibn SNe. The scaling is also approximately found from simulations of COs merging with red supergiants \citep{Schroder20}. The profiles are interestingly close to the CSM inferred from light curve modeling of Type Ibn SNe \citep{Maeda22}, shown in Figure \ref{fig:CSM_profile} as a shaded region\footnote{A possible caveat is that the velocity of the CBO $v_{\rm out}$ in our models at $\sim 10^{15}$ cm is a few 100 km s$^{-1}$, slower than $\sim 10^3$ km s$^{-1}$ typically inferred from narrow lines of SN Ibn \citep[e.g.,][]{Pastorello16}. To resolve this apparent tension (see also Section \ref{sec:2023fyq}), there has to be either (i) asymmetry effects, where $f_\Omega<1$ and line emission arises from the unimpeded (or less impeded) disk wind instead of the CBO, (ii) significant acceleration of the CSM by the SN radiation \citep[e.g.,][]{Tsuna23_CSMacc}, or (iii) spectral broadening by Thomson scattering \citep{Chugai01,Huang2018}.}. This supports binary interactions producing the CSM in at least some Type Ibn SNe, and motivates considering the terminal explosion in detail.

\subsection{Terminal Explosion of the Star: SN Ibn-like Explosion with/without an Early Bright Peak?}
\label{sec:explosion}

The final fates of these binaries undergoing unstable mass-transfer are likely to be diverse. The star may end its life as a core-collapse SN for some of our stellar models with higher masses, as envelope inflation ensues in the late burning stages at only months to years before core-collapse \citep{Wu22b}. In this case the mass transfer and precursor are suddenly truncated, and the light curve brightens to that of an interacting SN on a timescale of days to weeks.

Alternatively, the explosion of the star may happen as a result of the merger with the CO, in the case that the common envelope cannot be ejected \citep[e.g.,][]{Fryer98,Chevalier12,Soker13,Schroder20,Metzger22}. The energy of the explosion triggered by the merger is unclear due to large uncertainties in the accretion onto the CO, but can be comparable to typical SNe. For example, \cite{Metzger22} considers a merger-driven explosion scenario where the star is tidally disrupted at a separation comparable to its radius, accretes onto the CO at super-Eddington rates, and is expelled as a disk wind. We apply this to our binary models at the time of merger, where the donors have masses of $\approx 1.4$--$2.3~M_\odot$ with the majority confined in the carbon-oxygen core. The core has mass $M_{\rm core}\approx 1.4$--$1.5~M_\odot$ and radius $R_{\rm core}\approx$ (1--2)$\times 10^{-2}~R_\odot$. If we assume the whole core forms an accretion disk that circularizes at $R_{\rm core}$, time-integration of equation (\ref{eq:L_wind}) with $R_{\rm sph}=R_{\rm core}$ yields an energy budget of the disk wind of
\begin{eqnarray}
    E_{\rm wind}\sim 7\times 10^{51}\ {\rm erg}\left(\frac{M_{\rm NS}}{1.4M_\odot}\right)\left(\frac{M_{\rm core}}{1.5M_\odot}\right)\left(\frac{R_{\rm core}}{10^9\ {\rm cm}}\right)^{-1/2}.
\end{eqnarray}
where we used $p=0.5$. We note that this is likely an overestimation as the entire core may not accrete onto the CO. Some fraction can be unbound during the disruption process, and/or by substantial feedback from interaction with the disk wind \citep[e.g.,][]{Batta19}. In this conservative case we expect that the ``explosion energy" $E_{\rm exp}$ would not greatly exceed the initial binding energy of the core,
\begin{eqnarray}
    E_{\rm bind}&\approx&\frac{GM_{\rm core}(M_{\rm core}+M_{\rm NS})}{R_{\rm core}} \nonumber \\
    &\sim& 10^{51}\ {\rm erg}\left(\frac{M_{\rm core}}{1.5M_\odot}\right)\left(\frac{M_{\rm core}+M_{\rm NS}}{2.9M_\odot}\right)\left(\frac{R_{\rm core}}{10^9\ {\rm cm}}\right)^{-1}.
\end{eqnarray}
In either case, the disk wind can eject the bulk of the stellar material, both the core and any leftover He-rich envelope, with energies comparable to core-collapse SNe.

For both core-collapse and merger-driven explosions, the stellar ejecta from the explosion would run into the pre-existing CSM generated from the binary interaction, and convert part of its kinetic energy to radiation. Our models predict a total mass-loss up to merger of $M_{\rm CSM}\approx 0.1$--$0.3~M_\odot$, with a majority emitted in the ramp-up phase at the final 1--2 months before merger. This later ejected material has a typical velocity of $v_{\rm out}\sim 500$--$1000~{\rm km\ s^{-1}}$, and forms a confined, optically thick envelope with an extent of $R_{\rm env}\lesssim 10^{14}$ cm at the time of merger. 

For both explosion scenarios the ambient CSM at large scales ($\sim 10^{15}$ cm) are similar, but the merger case is expected to have an additional confined envelope of mass $M_{\rm env}\sim 0.1~M_\odot$. In this case, the ejecta is expected to first collide with this envelope, which power shock-cooling emission with timescale and luminosity \citep{Piro15} 
\begin{eqnarray}
    t_{\rm sc} &\approx & 2\ {\rm day} \left(\frac{\kappa_{\rm env}}{0.1\ {\rm cm^2\ g^{-1}}}\right)^{1/2}\left(\frac{E_{\rm exp}}{10^{51}\ {\rm erg}}\right)^{-1/4} \nonumber \\
    && \times \left(\frac{M_{\rm ej}}{2~M_\odot}\right)^{0.17} \left(\frac{M_{\rm env}}{0.1~M_\odot}\right)^{0.57},\\
    L_{\rm sc} &\approx&  1\times 10^{44}\ {\rm erg\ s^{-1}}\left(\frac{\kappa_{\rm env}}{0.1\ {\rm cm^2\ g^{-1}}}\right)^{-1}\left(\frac{E_{\rm exp}}{10^{51}\ {\rm erg}}\right) \nonumber \\
    && \times \left(\frac{R_{\rm env}}{6\times 10^{13}\ {\rm cm}}\right)\left(\frac{M_{\rm ej}}{2~M_\odot}\right)^{-0.7} \left(\frac{M_{\rm env}}{0.1~M_\odot}\right)^{-0.3},
\end{eqnarray}
where $\kappa_{\rm env}\approx 0.1$--$0.2\ {\rm cm^2\ g^{-1}}$ is the opacity of the expanding shock-heated envelope, and $M_{\rm ej}$ is the ejecta mass which corresponds to the total (core plus any leftover envelope) mass of the donor at merger for this scenario. This emission is observable as an early, bright peak separate from the emission powered by the later CSM interaction, and is a possibility to distinguish the two explosion scenarios. We expect the later CSM interaction (common for both merger and core-collapse) to generate a light curve similar to Type Ibn SNe, as the CSM profiles are similar to those inferred from light curve modeling of these SNe \citep{Maeda22}.

For the two explosion scenarios, the large variation of the CSM profile at inner radii ($r<10^{15}$ cm; Figure \ref{fig:CSM_profile}) can affect the propagation of the shock. This may also give some variation in later light curves and spectra of these Type Ibn-like transients, which is dependent on the properties of the ejecta ($M_{\rm ej}, E_{\rm exp}$) and can be investigated by detailed light curve modeling.

\section{Comparison to SN 2023fyq}
\label{sec:2023fyq}
\begin{figure}
    \centering
    \includegraphics[width=\linewidth]{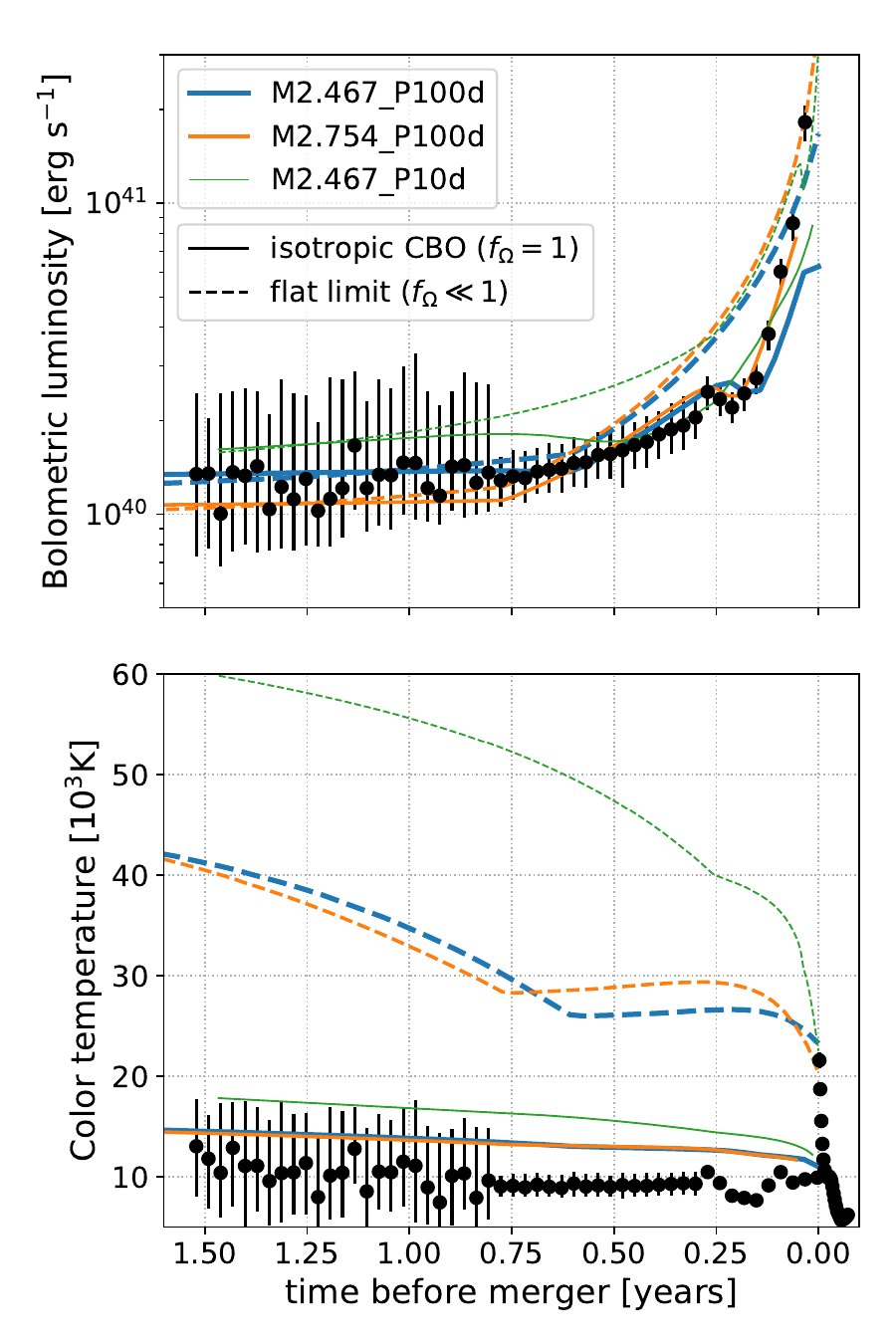}
    \caption{Comparison of our precursor model to the observed precursor of SN 2023fyq \citep{Dong24}. For the observations, we set $t=0$ in the plot as the first peak of the bolometric light curve. The observations are better reproduced with a CBO created from models with initial periods of 100 days, with a large covering fraction near unity.}
    \label{fig:lum_temp_withcbo_23fyq}
\end{figure}
We compare our model to the observations of SN 2023fyq \citep{Brennan24,Dong24}, which showed long-lasting precursor activity from several years, and displayed signatures of a Type Ibn SN as it continuously rose to peak. The observed luminosity and temperature are compared in Figure \ref{fig:lum_temp_withcbo_23fyq}. The models of initial period 100 days roughly reproduce the luminosity of the precursor from years before merger. The observed temperature ($\sim 10^4$ K) favors a significant covering fraction of the CBO ($f_{\Omega}$ close to unity) to emit in the optical.

We predict a light curve rise as observed, interestingly including the dip at $\approx 0.2$ yr before merger in the bolometric light curve. However in the final month(s) before merger when the light curve steeply rises, the model starts to become less reliable due to the assumptions of steady state. Therefore we are unable to robustly claim whether the sharp rise to peak is triggered by a merger or by a core-collapse SN. If the final explosion is triggered by a merger, the shock cooling emission estimated in Section \ref{sec:explosion} can potentially explain the timescale and luminosity around the peak (see also Figure 10 of \citealt{Dong24}). The mass range of the helium stars in our model partially overlaps with what is predicted to undergo mass ejection by vigorous off-center silicon burning \citep{Woosley19}, which may also contribute to the rise if the terminal explosion is due to core-collapse \citep{Dong24}. The CSM created by silicon flash can also power shock cooling emission \citep[e.g. Figure 15 of][]{Woosley19}, although the model lacks strong predictions for the CSM mass and radius.

The spectra taken during the precursor phase displays P-Cygni features in the He\textsc{i} lines of width $\approx 1700$ km s$^{-1}$. This velocity is similar to the disk wind in our models, but much faster than the CBO. Thus the CBO is likely not completely shielding the disk wind, and there are likely some solid angles where the wind is not significantly decelerated. Such an asymmetric structure is also supported by the multi-component emission line profiles observed at these times \citep{Brennan24}. To understand whether our setting can reproduce these features requires multi-dimensional spectral modeling with aspherical geometries for the wind and CBO, which is beyond the scope of this work.

The He\textsc{i} spectra at 37 and 17 days before explosion also display broad wings with widths of $\sim 14,000\ {\rm km\ s^{-1}}$, which seems too fast to explain by any precursor material. As discussed in \cite{Brennan24}, such wings can be produced via scattering by electrons in the CSM \citep[e.g.,][]{Chugai01,Huang2018,Ishii24}. We expect the spectra to be characterized by Thomson scattering from material outside the thermalization radius $r_{\rm therm}$. While the definition of a single $r_{\rm therm}$ is ambiguous for an asymmetric CSM, we find the color radius of our $f_{\Omega}=1$ model as $r_{\rm col}\approx 7\times 10^{13}$ cm, whereas blackbody fitting around these epoch finds a blackbody radius of $\approx (1$--$2)\times 10^{14}$ cm and temperature of $\approx 10^4$ K. The density profile of our CSM at these times approximately follows $\rho_{\rm CSM}(r)\approx 4\times 10^{-12}\ {\rm g\ cm^{-3}}(r/10^{14}{\rm cm})^{-3}$. The integrated scattering optical depth at $r\geq r_{\rm therm}$ is then
\begin{eqnarray}
    \tau_{\rm scat} &\sim& \frac{\kappa_{\rm scat}r_{\rm therm} \rho_{\rm CSM}(r=r_{\rm therm}) }{2} \nonumber \\
    &\sim& 20 \left(\frac{\kappa_{\rm scat}}{0.1\ {\rm cm^2\ g^{-1}}}\right) \left(\frac{r_{\rm therm}}{10^{14}\ {\rm cm}}\right)^{-2},
\end{eqnarray}
which is roughly consistent with the required $\tau_{\rm scat}\sim 20$ to reproduce the spectral width \citep{Brennan24}.

We roughly estimate the rate of such NS-Helium star binaries to consider whether they can explain the observation of SN 2023fyq. \cite{Wu22b} estimated that $\sim 10$\% of the Type Ibc SN progenitors have the appropriate mass to undergo late-stage mass transfer like our model. Given that $\sim 1/2$ of the time the donor helium star will be the less massive of the initial binary, and that $\sim 10\%$ of the binaries can survive upon the formation of the first NS \citep[e.g.,][]{Renzo19}, the rate of He-NS binary systems like considered here is expected to be $\sim 0.1\%$ of the core-collapse SN rate, which is $\sim 10$--$100\%$ of the SN Ibn rate \citep{Maeda22,Perley22}. This is roughly consistent with the finding of SN 2023fyq out of a few SN Ibn in a similar distance (SN 2006jc and SN 2015G; \citealt{Dong24}). However both the theoretical and observational estimates above are still highly uncertain, and both can be better constrained by detailed population synthesis calculations and/or future observations by e.g. Rubin/UVEX.

\section{Discussion and Conclusion}
\label{sec:conclusion}
We constructed a model for the observational appearance of unstable mass transfer from a SN progenitor star onto a CO companion. We specifically considered binaries of a NS and a low-mass ($2.5$--$3~M_\odot$) helium star, where the latter inflates and triggers intense mass transfer in the last years of its life \citep{Wu22b}.

Modeling the mass transfer using realistic helium star models, we found that the mass transfer rate (donor mass loss rate) starts at $|\dot{M}_*|\sim (10^{-3}$--$10^{-2})~M_\odot$ yr$^{-1}$, and rises gradually as the orbit slowly tightens. In the last months the mass-transfer rate sharply rises as mass transfer enters a runaway phase, leading to a rapid orbital shrinkage and merger of the binary.

The mass transfer process forms an outer CBO lost from the L2 point that carries most of the mass, and an inner accretion disk around the NS that launches a fast wind due to super-Eddington accretion (Figure \ref{fig:schematic}). We modeled emission powered by the disk wind launched from the NS, taking into account reprocessing in the disk wind and the CBO. We find that the disk wind, in concert with reprocessing by the CBO, powers slowly rising optical/UV transients of $10^{40}$--$10^{41}$ erg s$^{-1}$ for as long as several years. The emission temperature depends on the angle subtended by the CBO, with $\approx 10^4$ (a few $\times 10^4$) K for covering fractions of $f_\Omega\approx 1$ ($\ll 1$). We compared our model to a Type Ibn SN 2023fyq that showed a long precursor of several years, finding good agreement in both the luminosity and temperature of the precursor if the CBO covers a large fraction of the solid angle. SN 2023fyq is one of the closest Type Ibn SNe, and we expect future opical/UV surveys like Rubin and UVEX to find a large number of similar long-rising precursors of Type Ibn SNe.

If the accretor is not a CO, there will be no high-velocity disk wind. The precursor emission will then be much dimmer, likely undetectable for most extra-galactic transients. Our method could still be used to predict the density profile of the CSM produced during the transition to unstable mass transfer. While a Type Ibn SN may still be produced upon core-collapse, a bright precursor is not expected to be observed.

We conclude by mentioning the possible caveats and future avenues for our model. The model becomes less reliable as the binary approaches merger, due to the steady-state assumptions that break down when the orbital separation and mass loss rate rapidly changes in a timescale comparable to the orbital period. This assumption is problematic for both estimating the relative mass loss between the CBO and disk wind, as well as the luminosity and temperature of the transient. Hydrodynamical simulations of mass transfer, including viscous heating that trigger L2 mass-loss, would be required to reliably estimate the emission at the steeply rising phase.

We have chosen to model the orbital evolution by \texttt{RLOF} rather than stellar evolution codes like MESA, as the former can stably solve the evolution until merger. While the orbital evolution is calibrated from numerical simulations \cite{Macleod20}, the evolution of the donor radius is simplified as discussed in Section \ref{sec:rlof_model}. Variations in radius, which are captured in binary modeling including stellar evolution \citep[e.g.,][]{Dewi03,Wu22b}, can cause variations in the rate of mass transfer. Though the precursor luminosity is only weakly sensitive to the mass transfer rate (top panels of Figure \ref{fig:disk_wind_properties}), changes in the mass transfer rate would affect the degree of reprocessing by the CBO, and can alter the temperature of the precursor emission.

While the accretion physics in our model is simple, it sensitively depends on some of the parameters. Increasing the mass of the CO will greatly raise the precursor luminosity as it increases the accretion rate onto the CO \citep{Lu23}, and the wind energy budget is proportional to both the accretion rate and the CO mass. The power-law index $p$ for the accretion rate (Section \ref{sec:wind}) is also important, and changing from our fiducial $p=0.5$ to $p=0.4$ or $p=0.6$ \citep{Yuan12,Hu22} alters the luminosity by a factor of a few. However, neither of these parameters greatly changes the light curve morphology, of a steady phase followed by a steep rise.

Our formalism can be straightforwardly applied to other transients that show similar long-rising precursors. For example, a Type IIn SN 2021qqp showed a precursor for at least 1 year \citep{Hiramatsu24}, with light curve morphology similar to what we predict. The rising timescale of $\lesssim 100$ days is similar to our models, but the luminosity ($\approx -15$ mag in r-band at 300 days before explosion) is 10--100 times brighter. In our model the much brighter luminosity requires larger power from the inner disk wind, which may be achieved by a heavier (i.e. BH) companion.

\cite{Tsuna24} proposed a mechanism for luminous SN precursors, where the CSM is supplied by an outburst instead of continuous mass transfer (this work). Our model struggles to explain the precursors that do not have a smooth rise, such as those of SN 2009ip \citep[e.g.,][]{Pastorello13} as well as other SN Ibn like SN 2006jc and 2019uo \citep{Pastorello07,Strotjohann21}. For those precursors outburst(s) like considered in \cite{Tsuna24} is a more likely scenario. The key advantage of our model with continuous energy injection is that it can explain extremely long-duration precursors like SN 2023fyq. Year-long precursors are difficult to explain by the model of \cite{Tsuna24}, as the timescale in their model is limited by the diffusion timescale of the ejected CSM for helium stars (days to months; see their Figure 4).

We did not fully explore the vast parameter space expected for various donor stars, accretor mass, or compositions. We thus release our source code\footnote{\url{https://github.com/DTsuna/merger_precursor.git}}, which enables one to construct light curves given the binary parameters and a time evolution of $\dot{M}_*$. While there are theoretical uncertainties in the mass transfer history, one plausible model is to assume a power-law phase followed by a rise to a singularity at $t=t_0$ \citep{Webbink77,Pejcha14}
\begin{eqnarray}
    \dot{M}_*(t)=-\dot{M}_0\left(\frac{t_0}{t_0-t}\right)^{\delta}.
\end{eqnarray}
The evolution of $\dot{M}_*$ in our models indeed show an evolution consistent with this assumption, and such formulation reasonably explains the observed evolution of luminous red novae like V 1309 Sco \citep{Pejcha17}. Another potential application is to couple this model with multi-dimensional hydrodynamical simulations of unstable mass transfer, which can capture the evolution of the donor structure and angular structure of the CBO that are simplified in this work.

\begin{acknowledgments}
We thank Wenbin Lu for discussions on the opacity modeling, and the anonymous referee for helpful comments that improved this manuscript. D. T. is supported by the Sherman Fairchild Postdoctoral Fellowship at the California Institute of Technology. S.C.W. is supported by the National Science Foundation Graduate Research Fellowship under grant No. DGE-1745301. J.F. is grateful for support from the NSF through grant AST-2205974. Research by Y.D. is supported by NSF grant AST-2008108.
\end{acknowledgments}

\appendix
\section{Modeling the Absorption Opacity}
\label{sec:app_opacity}
For a given $\rho$ and $T$, we calculate the absorption opacity as $\kappa_{\rm abs}=\kappa_{\rm R}-\kappa_{\rm scat}$, where we obtain the total (Rosseland-mean) opacity $\kappa_{\rm R}$ from tabulated opacities available in MESA, and the scattering opacity $\kappa_{\rm scat}=n_e\sigma_{\rm Thom}/\rho$ by solving the electron number density $n_e$ from Saha equations (with $\sigma_{\rm Thom}$ being the Thomson cross section). The Saha equations are solved using a code in the open-source light curve modeling tool CHIPS \citep{Takei24}, which calculates the ionization of helium, carbon and oxygen for a given $\rho$ and $T$ and was developed for the purpose of light curve modeling of interaction-powered SNe. The latter two elements are solved up to CV and OV, and ions of higher levels are neglected for simplicity. 

The opacity tables are split by temperature into OPAL \citep{Iglesias96} and low-temperature tables \citep{Ferguson05}, with a small temperature overlap at around $10^4$ K. However for a He-dominated composition we found large discrepancies between these two tables at low densities, which may be due to OPAL neglecting line absorption by neutral helium \citep[Section 4.3 of][]{Ferguson05}. We have thus adopted for $\kappa_{\rm R}$ in the overlapping region the values in the low-temperature opacity tables, and neglected the OPAL data there. 

For densities too low that are outside what are covered by the tables (i.e. $\rho < \rho_{\rm min}=10^{-8}(T/10^6\ {\rm K})^3\ {\rm g\ cm^{-3}}$), we obtain the absorption opacity by using the value at $\rho=\rho_{\rm min}$ (i.e. assuming $\kappa_{\rm abs}$ is independent of $\rho$), and obtain $\kappa_{\rm R}$ by adding this $\kappa_{\rm abs}$ to $\kappa_{\rm scat}$. We expect the bound-bound opacity to be nearly independent of $\rho$ (although ionization states depend slightly on $\rho$), and the bound-free/free-free opacity to be proportional to $\rho$. For the densities of our interest which are much lower than stellar interiors, we expect the former to dominate over the latter. For low temperatures of $T\lesssim 2000$ K, the \cite{Ferguson05} table shows large contributions from dust grains ($\kappa_{\rm abs}\sim 10^{-2}$--$10^{-1}$ cm$^2$ g$^{-1}$). However, dust in the vicinity of the system is likely sublimated by the precursor emission, and we thus neglect $\kappa_{\rm abs}$ below $2000$ K. Copious dust may form closer to merger, when the CBO with large mass loss rate has expanded to a large radius.

\begin{figure*}
   \centering
   \includegraphics[width=\linewidth]{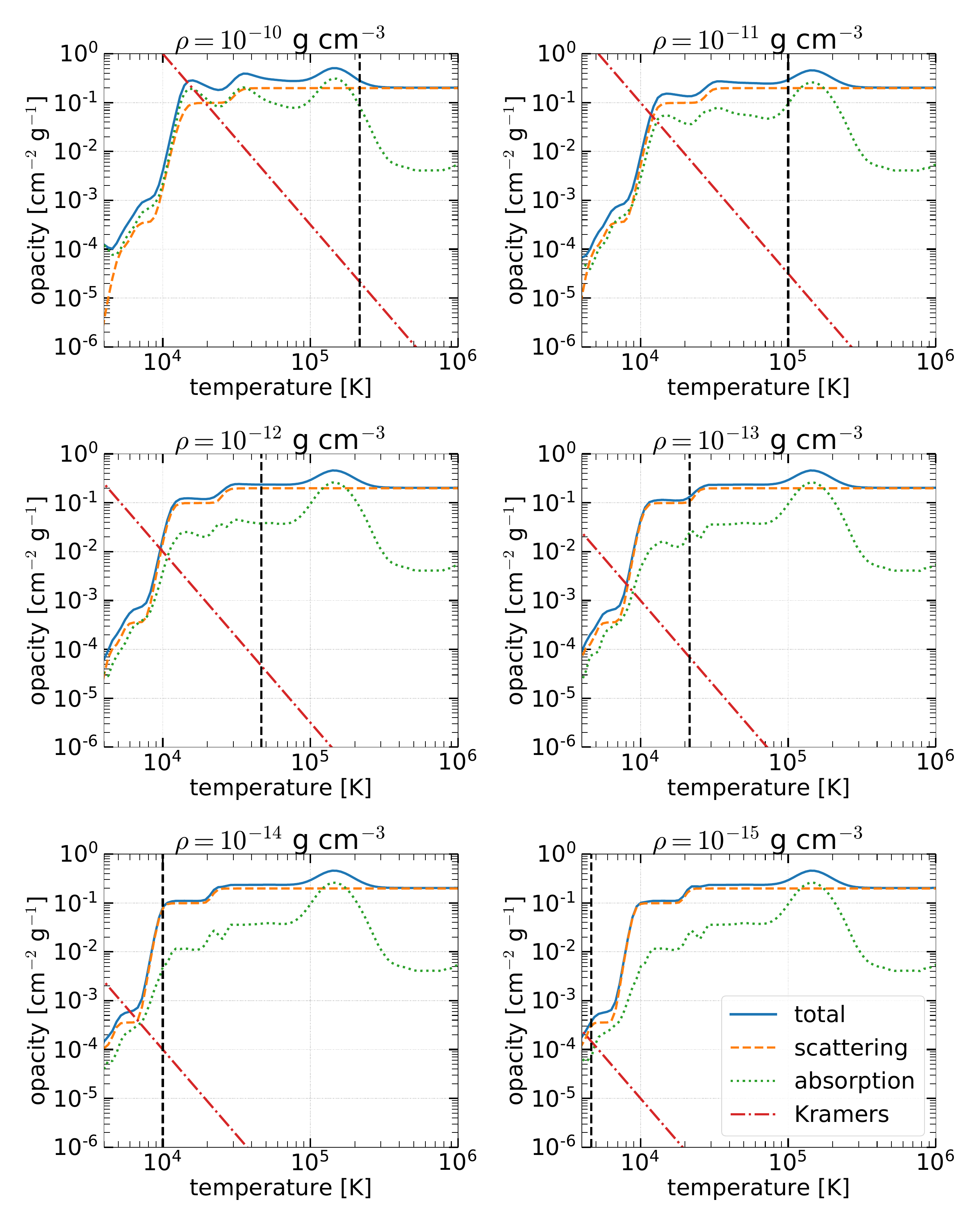}
    \caption{Opacities used in this model, for representative densities of $\rho=10^{-10}$--$10^{-15}\ {\rm g\ cm^{-3}}$. Blue solid, orange dashed, and green dotted lines in each plot correspond to the total (scattering + absorption), scattering and absorption opacities respectively. The dash-dotted straight lines indicate the Kramer's opacity law, and the black vertical lines indicate the temperatures above which we rely on extrapolation from the table.}
    \label{fig:opacities}
\end{figure*}

Figure \ref{fig:opacities} shows the opacities for a range of densities $\rho=10^{-15}$--$10^{-10}\ {\rm g\ cm^{-3}}$ relevant in this work. Independent of the density, both the scattering and absorption opacities steeply drop at $T\approx 10^4$ K due to helium recombination. Our model is approximate and may be inaccurate at high temperatures of $T\gtrsim$ (a few) $\times 10^5$ K where ionizations beyond CV and OV occur, and densities of $\rho\lesssim 10^{-13}\ {\rm g\ cm^{-3}}$ where we rely on extrapolation from the table at the most interesting temperature range $T\lesssim$ (a few)$\times 10^4$ K. Nevertheless, using these absorption opacities (rather than simpler prescriptions like the Kramer's law) is important for observational predictions in the optical and near-UV, as it correctly captures the behaviour of the opacity dropping by orders of magnitude at $T\lesssim 10^4$ K.

\bibliography{references}
\bibliographystyle{aasjournal}

\end{document}